\documentclass[conference]{IEEEtran}
\usepackage{amssymb,amsmath,amsthm,bm}
\usepackage{graphicx,color}
\usepackage{subfigure}
\graphicspath{{figures-pdf/}}
\usepackage{cite}
\usepackage{url}

\usepackage{titlesec}
\titlespacing{\section}{0pt}{5pt}{2pt}
\titlespacing{\subsection}{0pt}{3pt}{2pt}
\titlespacing{\subsubsection}{0pt}{1.5pt}{1pt}

\usepackage[mathlines]{lineno}

\usepackage{multirow}
\usepackage{setspace}

\DeclareMathOperator\round{round}
\DeclareMathOperator\rank{rank}
\makeatletter
\newcommand{\dotminus}{\mathbin{\text{\@dotminus}}}
\newcommand{\@dotminus}{
  \ooalign{\hidewidth\raise1ex\hbox{.}\hidewidth\cr$\m@th-$\cr}
}
\makeatother

\newlength\imagewidth
\newlength\imagewidtha
\newlength\imagewidthb
\newlength\figwidth
\newlength\figwidtha
\newlength\figsep
\ifCLASSINFOpdf
\else
\fi
\begin{document}
%
\title{Joint Quantization and Diffusion for Compressed Sensing Measurements of Natural Images}

\author{
\IEEEauthorblockN{Leo Yu Zhang \\ and Kwok-Wo Wong}
\IEEEauthorblockA{Department of Electronic Engineering, \\City University of Hong Kong, Hong Kong\\
Email:leocityu@gmail.com \\ and itkwwong@cityu.edu.hk}
\and
\IEEEauthorblockN{Yushu Zhang}
\IEEEauthorblockA{School of Electronics \\ and Information Engineering,\\ Southwest University,\\ Chongqing 400715, China\\
Email: yushuboshi@163.com}
\and
\IEEEauthorblockN{Qiuzhen Lin}
\IEEEauthorblockA{College of Computer Science\\ and Software Engineering, \\Shenzhen University, \\Shenzhen 518060, China \\
Email:  qiuzhlin@szu.edu.cn}
}


%


\maketitle

\begin{abstract}
Recent research advances have revealed the computational secrecy of the compressed sensing (CS) paradigm.
Perfect secrecy can also be achieved by normalizing the CS measurement vector. However, these findings are established on real measurements 
while digital devices can only store measurements at a finite precision.
Based on the distribution of measurements of natural images sensed by structurally random ensemble, a joint quantization and diffusion approach
is proposed for these real-valued measurements. In this way, a nonlinear cryptographic diffusion is intrinsically imposed on the
CS process and the overall security level is thus enhanced.
Security analyses show that the proposed scheme is able to resist known-plaintext attack while the original CS scheme without quantization cannot.
Experimental results demonstrate that the reconstruction quality of our scheme is comparable to that of the original one.
\end{abstract}


%
\IEEEpeerreviewmaketitle

\section{Introduction}
Compressed sensing (CS) has received much research attention in the past decade \cite{Tao:BPunderRIP:TIT05,Donoho:CSIntro:TIT06,Richard:CSIntro:SPM07}
since it samples a sparse signal at a rate lower than that required by the traditional sampling theorem.
For a signal $\mathbf{x}$ which can be represented by $k$ out of $N$ significant components under a certain transformation $\mathbf{\Psi}$,
the CS theory states that $\mathbf{x}$ can be faithfully recovered from its compressive sampled linear measurements
$\mathbf{y} = \mathbf{\Phi} \cdot \mathbf{x}$ under the condition that the sensing matrix $\mathbf{A} = \mathbf{\Phi} \cdot \mathbf{\Psi}$ satisfies the restricted isometry
property (RIP) \cite{Richard:ProveRIP:2008}.

In the CS process, Gaussian or Bernoulli matrices are commonly employed as the measurement matrix due to their universality
and optimal measurement performance. When CS is used to sample natural images,
new issues, such as fast computation and efficient storage, arise.
Structurally random matrices (SRM) are potential candidates to meet these new requirements \cite{Do:TSP:SRM12}.
In general, CS employing SRM can be formulated as
\begin{equation}
\label{eq:SRMstructure}
\mathbf{\Phi} = \sqrt{ \frac{N}{M}} \mathbf{DFR},
\end{equation}
where $\mathbf{R}$ is either a random permutation matrix or a diagonal matrix whose diagonal elements are independent and identically distributed (i.i.d.) Bernoulli variables, $\mathbf{F}$ is an
orthogonal matrix which can be chosen from fast transformations such as DFT, DCT or WHT, and $\mathbf{D}$ is a down sampling operator
which selects $M$ measurements out of $N$ ones uniformly. For the purpose of efficient storage, $\mathbf{F}$ can be in the form of a block diagonal matrix, i.e.,
\begin{equation}
\mathbf{F} = \left(
\begin{array}{cccc}
\mathbf{{F}}_b & 0 & \ldots & 0\\
0 & \mathbf{{F}}_b & \ldots & 0\\
\vdots & \vdots & \ddots  &\vdots \\
0 &0  & \ldots & \mathbf{{F}}_b
\end{array} \right),
\label{eq:blockmatrix}
\end{equation}
where $\mathbf{{F}}_b$ is an orthogonal matrix of size $B\times B$.
In summary, the compressed sensing process using SRM can be carried out in three steps: pre-randomization of the signal to be sensed;
orthogonal transformation and down sampling.

When the measurement matrix $\mathbf{\Phi}$ is treated as a secret key, it is reported in \cite{Candes:CS:TIT06} that unauthorized
users cannot decode the measurement vector because they do not know which random subspace the
measurement vector is expressed. Thus CS can be considered as possessing the property of a symmetric cipher.
Rachlin \textit{et al}. proved that breaking a CS-based cipher by systematically searching in the key space
is computational infeasible \cite{Rachlin:secrecy:08}. This finding was later extended to Shannon's perfect secrecy
by Bianchi \textit{et al}. through normalizing the ciphertext \cite{Bianchi:SecRLM:ICASSP14}.

However, the application of these findings in practical scenarios is hampered since
a) CS-based ciphers require one-time-pad keys which lead to key delivery and management problems;
b) the achieved secrecy is based on real-valued measurements but digital devices can only store
finite-precision numbers. These motivate us to investigate the security characteristics
of CS with quantized measurements. Making use of the distribution of measurements of natural images, we suggest quantizing
the measurement vector to integers in the interval $[0,255]$ using a first-order $\Sigma \Delta$
quantizer \cite{Feng2014SPLQuanti}. The similarities between error feedback mechanism of the quantizer and cryptographic diffusion primitive inspire us to
design a joint quantization and diffusion approach for the measurement vector. With the intrinsic diffusion property, the resistance to known-plaintext attack (KPA)
of CS-based cipher is thus improved. Simulation results justify that the reconstruction quality of the proposed
approach is comparable to the original CS without quantization. To the best of our knowledge, this is the first study of embedding
security measures in the quantization process of CS.

The rest of this paper is organized as follows. Sec.~\ref{sec:sampling} is an introduction of our joint quantization and diffusion approach for CS measurements.
Sec.~\ref{sec:securesampling} presents a secure and fast CS scheme for natural images based on the proposed approach.
Security analyses and simulation results can be found in Sec.~\ref{sec:analysis} while other possible applications of the proposed scheme
are discussed in Sec.~\ref{sec:extension}.
The last section concludes our work.

\section{A Joint Quantization and Diffusion Approach for Measurements of Natural Images}
\label{sec:sampling}
\begin{figure}[!htb]
\centering
\begin{minipage}[t]{\imagewidth}
\centering
\includegraphics[width=\imagewidth]{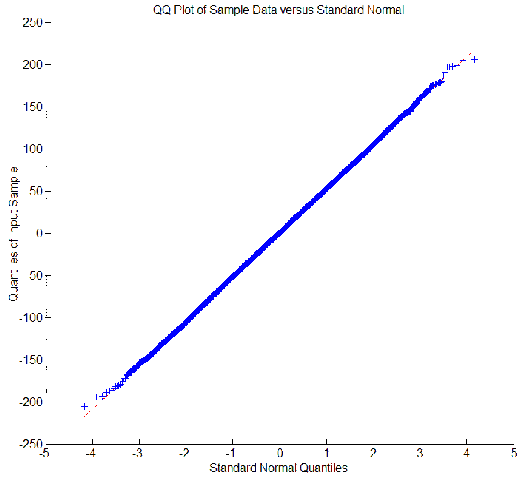}
\end{minipage}
\centering \caption{QQ plot of measurements obtained using structurally random DCT matrix with block size $B=32$ and sampling rate $SR =M/N= 0.5$. The test image
is $256 \times 256$ \textit{Lena} image.}
\label{fig:qqplot}
\end{figure}

\begin{figure*}[hth]
\centering
\subfigure[]{
\begin{minipage}[t]{0.2\linewidth}
\centering \includegraphics[width=\imagewidtha]{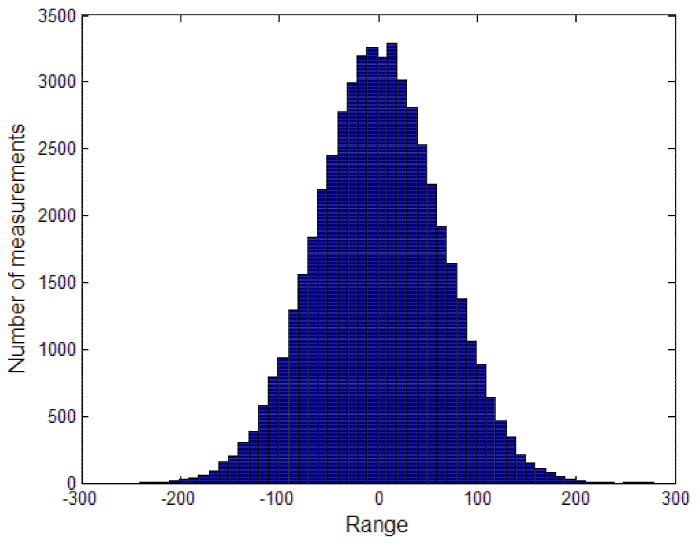}
\end{minipage}}
\subfigure[]{
\begin{minipage}[t]{0.2\linewidth}
\centering \includegraphics[width=\imagewidtha]{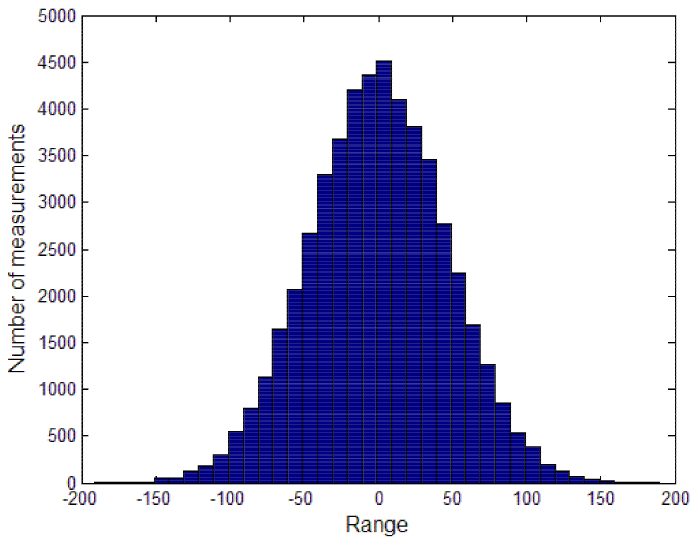}
\end{minipage}}
\subfigure[]{
\begin{minipage}[t]{0.2\linewidth}
\centering \includegraphics[width=\imagewidtha]{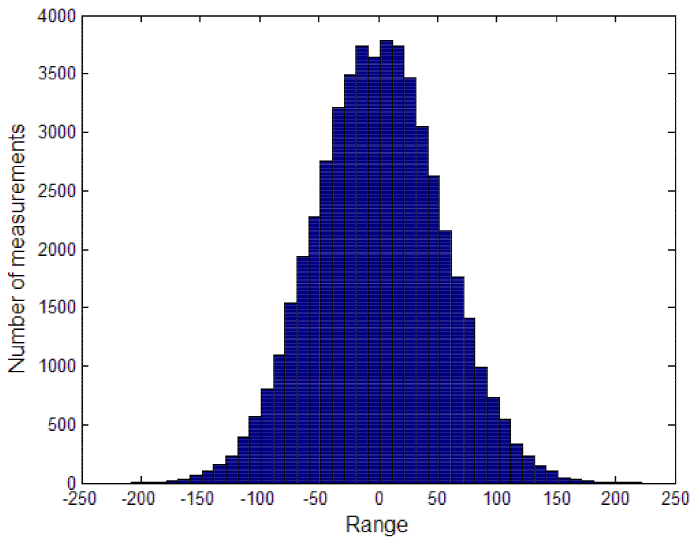}
\end{minipage}}%
\subfigure[]{
\begin{minipage}[t]{0.2\linewidth}
\centering \includegraphics[width=\imagewidtha]{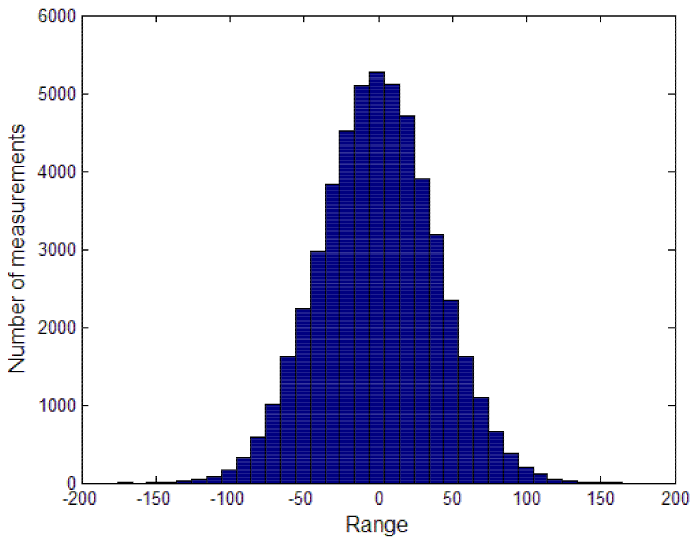}
\end{minipage}}
\caption{Distribution of measurements obtained from sampling several test images at different settings ($SR=0.8$):
(a) \textit{Cameraman}, DCT with block size $B=32$;
(b) \textit{House}, DCT with block size $B=64$;
(c) \textit{Peppers}, WHT with block size $B=128$;
(d) \textit{Baboon}, DFT with block size $B=256$.}
\label{fig:distribution}
\end{figure*}

As observed from Eqs.~(\ref{eq:SRMstructure}) and (\ref{eq:blockmatrix}), each measurement can be considered as a weighted sum of $B$ random pixels.
According to the Central Limit Theorem (CLT), the samples follow the Gaussian distribution asymptotically.
To explain this phenomenon in more detail, we carry out some experiments using different block sizes and orthogonal transformations.
Fig.~\ref{fig:qqplot} is a plot of the quantities of input samples vs that of standard normal, using \textit{Lena} as the test image. It shows that the measurements
obtained by sensing a natural image using the structurally random DCT matrix behave like Gaussian random variables.
More simulations have been performed to sample other test images using different transform matrices and block sizes. The distributions of the measurements
are depicted in Figs.~\ref{fig:distribution}(a)-(d).
\vspace*{-0.6\baselineskip}
\vspace*{-0.4\baselineskip}

\vspace*{-0.4\baselineskip}

Inspired by the $3$-sigma rule of Gaussian variables, i.e., about $99.7\%$ of values drawn from a Gaussian distribution are within three standard deviations from the mean,
we propose to quantize these measurements to the range $[0, 255]$ using a first-order $\Sigma \Delta$ quantizer.
The quantization process can be expressed as
\vspace*{-0.5\baselineskip}
\begin{eqnarray}
\label{qe:sigmadelta}
q_i &=&\mathcal{Q}(u_{i-1}+y_i),  \\
u_i &=& \left\{
\begin{array}{rl}
(u_{i-1} + y_i) -q_i & \text{if } 0< q_i < 255 ,\\
u_{i-1} & \text{otherwise},
\end{array} \right.
\label{qe:errorcal}
\end{eqnarray}
where $u_{0}$ is the initial quantization error, $\mathbf{q}= \{q_i\}_{i=1}^{M}$ is the quantized output of the measurement vector $\mathbf{y}=\{y_i\}_{i=1}^{M}$ and
$ \mathcal{Q}(\cdot)$ is a $8$-bit quantizer governed by
\vspace*{-0.3\baselineskip}
\begin{equation*}
\label{eq:uniformquantization}
\mathcal{Q}(a) = \left\{
\begin{array}{rl}
\round(a)+127 & \text{if } -127.5 \leq a < 128.5 ,\\
0 & \text{if } a<-127.5, \\
255 & \text{if } a \geq 128.5,
\end{array} \right.
\end{equation*}
where $\round(a)$ returns the nearest integer of $a$.

From the above description, it is easy to find that the quantization error
at the $i$-th step is updated by the error at the previous step using Eq.~(\ref{qe:errorcal}).
This updated error will influence the quantization process of the next element $y_{i+1}$ of the measurement vector, as governed by Eq.~(\ref{qe:sigmadelta}).
In this way, the $i$-th quantized value $q_i$ is affected by the initial error $u_0$ through an accumulated manner,
which possesses certain intrinsic similarity with the cryptographic diffusion primitive.
Suppose that a key stream $\mathbf{V} = \{v_i\}_{i=0}^{M}$ has been generated, a typical diffusion approach using nonlinear modular addition
can be expressed as
\vspace*{-0.4\baselineskip}
\begin{equation}
q_i^* = q_i \dotplus v_i \dotplus q_{i-1}^*,
\label{eq:diffusion}
\end{equation}
where $q_0^* = v_0$ is the initial value and $a\dotplus b = (a+b) \bmod 256$. With the help of the value $q_{i-1}^*$ and the key stream,
the $i$-th quantized measurement $q_i$ can be decrypted by the following reverse operations:
\vspace*{-0.4\baselineskip}
\begin{equation}
q_i = q_i^* \dotminus v_i \dotminus q_{i-1}^*,
\label{eq:inversediffusion}
\end{equation}
where $a \dotminus b = (a-b+256) \bmod 256$. As a result, a joint quantization and diffusion approach for the CS measurements can be derived from
Eqs.~(\ref{qe:sigmadelta}), (\ref{qe:errorcal}), (\ref{eq:diffusion}) and (\ref{eq:inversediffusion}). The process can be expressed as
\vspace*{-0.4\baselineskip}
\begin{eqnarray*}
q_i^* &=&  \mathcal{Q}(u_{i-1}+y_i) \dotplus v_i \dotplus q_{i-1}^*, \\ \nonumber
u_i   &=&
 \left\{
\begin{array}{rl}
(u_{i-1} + y_i) - q_i & \text{if } 0< q_i < 255 ,\\
u_{i-1} & \text{otherwise},
\end{array} \right.
\end{eqnarray*}
It should be noticed that $q_i^*$ has a uniform distribution rather than a Guassian one, as caused by the randomness of the key
stream $\{v_i\}_{i=0}^{M}$. Moreover, the quantization error made in the previous step still accounts for the quantization of the next measurement.
Thus, all the advantages of the original $\Sigma \Delta$ quantization method \cite{Feng2014SPLQuanti} are reserved.

\section{~Fast Compressed Sensing of Natural Images with Secrecy}
\label{sec:securesampling}
Before describing the details of a fast CS scheme with secrecy for natural images, we first address the problems of constructing the
permutation matrix $\mathbf{R}$ and the down sampling operator $\mathbf{D}$ to form the SRM as well as generating the
key stream $\mathbf{V}$ for quantization and diffusion.
Without loss of generality, we assume that a cryptographic random number generator (RNG) which can produce random numbers uniformly in the interval $[1, N]$
is available.
As stated in \cite[Sec. 5.3]{Cormen:IntroductionToAlg:09}, a random permutation sequence $\{\tau(i)\}_{i=1}^{N}$ of the set $\{1,2,3,\cdots, N-1, N\}$ can be generated at a complexity of $O(N)$ by the following method:
\vspace*{-0.4\baselineskip}
\begin{itemize}
\item{Step 1.} Initialize $\tau(i) = i$ for all $i \in \{1,2,\cdots, N\}$.
\item{Step 2.} Set $i=1$.
\item{Step 3.} Generate a random number $j\in [i, N]$ using the RNG.
\item{Step 4.} Swap $\tau(i)$ and $\tau(j)$.
\item{Step 5.} If $i<N$, let $i=i+1$ and turn to Step~3; otherwise, return $\{\tau(i)\}_{i=1}^{N}$.
\end{itemize}

Given a $128$-bit secret key $\mathbb{K}$ as the seed of the RNG, two sequences
$\{\tau(i)\}_{i=1}^{N}$ and $\{\tau(j)\}_{j=1}^{M}$ are produced by running the RNG for $(N+M)$
times from the seed. Then the permutation matrix $\mathbf{R}$ and the down sampling matrix $\mathbf{D}$ are
determined by\footnote{When $\mathbf{R}$ is a diagonal matrix
whose diagonal elements are i.i.d. Bernoulli variables, we can set $\mathbf{r}_i = (0,0, \cdots, (-1)^{\tau(i) \bmod 2}_i, 0, \cdots, 0)$ instead.}
\vspace*{-0.4\baselineskip}
\begin{eqnarray*}
\mathbf{r}_i &= (0,0, \cdots, 1_{\tau(i)}, 0, \cdots, 0),  \\ \nonumber
\mathbf{d}_j &= (0,0, \cdots, 1_{\tau(j)}, 0, \cdots, 0),\nonumber
\end{eqnarray*}
where $ i \in [1,N]$, $j \in [1,M]$, $\mathbf{r}_i$ and $\mathbf{d}_j$ denote the $i$-th row of $\mathbf{R}$ and the $j$-th row of $\mathbf{D}$, respectively.
Similarly, run the RNG for a further of $(M+1)$ times and produce $\{\tau(k)\}_{k=1}^{M+1}$, the key stream $\mathbf{V} = \{v_i\}_{i=0}^{M}$  for diffusing the measurements
is obtained by setting $v_i = \tau(i+1) \bmod 256$.

Making use of the above notations, the proposed fast CS method with secrecy for a $H \times W~(N=H \times W)$ natural image $\mathbf{P}$ contains the following operation steps:
\begin{itemize}
\item{\textit{Initialization}:}
Construct $\mathbf{R}$, $\mathbf{D}$ and $\mathbf{V}$ from the key $\mathbb{K}$ as described above, then stretch $\mathbf{P}$ to a vector
$\mathbf{x}$ by stacking its columns.
\item{\textit{Pre-randomization}:} Randomize $\mathbf{x}$ using the permutation matrix $\mathbf{R}$.
\item{\textit{Orthogonal Transformation}:} Apply a chosen fast transform, such as DCT or DFT, to the output of the previous step.
\item{\textit{Down Sampling and Normalization}:} Select $M$ real measurements $\mathbf{y} = \{y_i\}_{i=1}^{M}$ out of $N$ ones uniformly using $\mathbf{D}$ and normalize the result.
\item{\textit{Quantization and Diffusion}:} Apply the joint quantization and diffusion approach to $\mathbf{y}$ with the help of the key stream $\mathbf{V}$.
\end{itemize}
The whole scheme can be expressed as
\vspace*{-0.4\baselineskip}
\begin{equation*}
\mathbf{q}^* = \mathcal{QD}( \mathbf{\Phi} \cdot \mathbf{x} ) = \mathcal{QD}(  \sqrt{ \frac{N}{M}} \mathbf{DFR} \cdot \mathbf{x} ),
\label{eq:summarize}
\end{equation*}
where $\mathcal{QD}(\cdot)$ is the joint quantization and diffusion operation described in Sec.~\ref{sec:sampling}.

To reconstruct the original image $\mathbf{P}$, or equivalently $\mathbf{x}$, one should first obtain the quantization vector $\mathbf{q}$ from
the inverse diffusion equation (\ref{eq:inversediffusion}) using the key stream $\mathbf{V}$. Then, the measurement matrix
used for sampling is exactly retrieved
by the decoder using $\mathbb{K}$. Finally, the gradient projection for sparse reconstruction (GPSR) algorithm \cite{figueiredo2007gradient} is employed to tackle the following
optimization problem\footnote{A simple trick is employed to handle the saturated quantized measurements. We will introduce it in Sec.~\ref{subsec:reconstruction}.}
\vspace*{-0.4\baselineskip}
\begin{equation}
\label{eq:l1opt}
  \min \|\mathbf{{s}}\|_1~~\text{subject to } \| \mathbf{q} -\mathbf{\Phi \Psi}  \mathbf{{s}}\|_2^2 < \varepsilon,
\end{equation}
and an approximate version of $\mathbf{x}$ can be determined by $\mathbf{\hat{x}} = \mathbf{\Psi}\cdot \mathbf{s}$.

\section{Security Analyses and Numerical Simulations}
\label{sec:analysis}

\subsection{Known-Plaintext Attack}
It is widely recognized that the sensitivity to changes in plaintext plays a significant role in determining a cipher's security level, especially its
resistance to plaintext attacks. The encryption process of a CS-based cipher without quantization is based on random projection, which is linear.
Therefore, it is not surprise that the ciphertexts or measurements are not sensitive to the changes in plaintext.
Based on this observation, an attacker can carry out known-plaintext attack on the non-quantized CS, which can be expressed as
\begin{equation*}
\mathbf{y} =  \mathbf{\Phi} \cdot \mathbf{x}  =  \mathbf{\Phi \Psi} \cdot \mathbf{s}  .
\label{eq:originalsampling}
\end{equation*}
The procedures of the attack are described as follows:
\begin{itemize}
\item Check whether $n>>N$, where $n$ denotes the number of known plaintext and ciphertext pairs the attacker possesses.
\item Randomly choose a natural image $\mathbf{P}^j$ whose vectorized version $\mathbf{x}^j$ is linearly independent of all the $j-1$ existing known plain-images.
\item Collect all the selected known-images and the corresponding ciphertexts if $j=N$. Then denote $\mathbf{X}= [\mathbf{x}^1, \mathbf{x}^2, \cdots, \mathbf{x}^N]$ and
$\mathbf{Y}= [\mathbf{y}^1, \mathbf{y}^2, \cdots, \mathbf{y}^N]$.
\end{itemize}
Now, it is clear that $\rank(\mathbf{X}) =N$ and an approximate version of $\mathbf{\Phi}$ can be obtained by $\mathbf{\hat{\Phi}} = \mathbf{Y} \cdot \mathbf{X}^{-1}$.

In the CS process employing our joint quantization and diffusion, the original ciphertext $\mathbf{y}$, which is linearly dependent on the plain-image,
is converted to a quantized and diffused vector $\mathbf{q}^*$, which is uniformly distributed in $[0, 255]$. Thus, the linear relationship between the plain-image and the ciphertext
is disturbed and the KPA attack described above becomes ineffective.

\subsection{Reconstruction}
\label{subsec:reconstruction}
Different from the uniform quantization proposed in \cite{Liu14SRBitDepth}, we map the measurements whose values are outside the interval $(-127.5, 128.5]$ to $0$ or $255$, as described in Sec.~\ref{sec:sampling}.
Taking this effect into consideration, we simply reject both the positively and the negatively saturated measurements during reconstruction.
In this way, the optimization problem (\ref{eq:l1opt}) becomes
\vspace*{-0.4\baselineskip}
\begin{equation*}
\label{eq:reformedl1opt}
  \min \|\mathbf{{s}}\|_1~~\text{subject to } \| \mathbf{q} -\mathbf{\Phi' \Psi}  \mathbf{{s}}\|_2^2 < \varepsilon,
\end{equation*}
where $\mathbf{\Phi}'$ is composed of all the rows of $\mathbf{\Phi}$ with quantized measurements $q_i$ satisfying $0<q_i<255$.

\begin{figure}[!htb]
\centering
\begin{minipage}[t]{\imagewidth}
\centering
\includegraphics[width=\imagewidth]{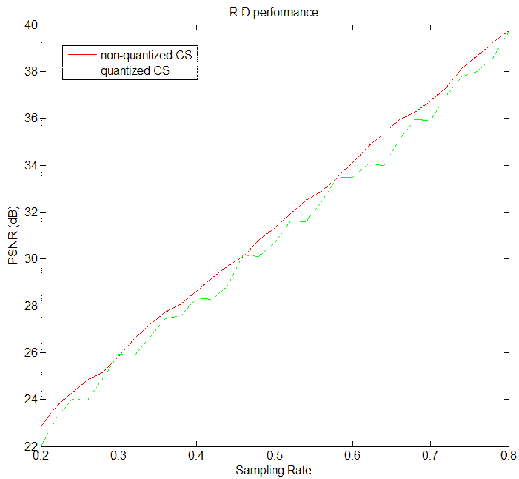}
\end{minipage}
\centering \caption{Reconstruction performance of the quantized and the non-quantized CS using SRM.}
\label{fig:RDcurve}
\end{figure}

The performance of the joint quantization and diffusion approach for CS measurements is depicted in Fig.~\ref{fig:RDcurve}, where the
$256\times 256$ \textit{Lena} image is compressively sampled by the block DCT matrix at a block size $B=32$. The benchmark for comparison is non-quantized CS using SRM
under the same setting \cite{Do:TSP:SRM12}. It is clear in Fig.~\ref{fig:RDcurve} that the performance gap between the quantized and the non-quantized CS is acceptable and the average loss
in peak signal-to-noise ratio (PSNR) due to quantization is around $0.6$dB.

\vspace*{-0.4\baselineskip}

\section{Extension to Other Applications}
\label{sec:extension}
We have studied the security characteristics of the proposed CS process with quantization by treating it as a symmetric cipher. Its feasibility for compressing encrypted images in secure signal processing application
\cite{kumar2009lossy,zhang2011compressing} will be explored here.

\begin{figure}[!htb]
\centering
\begin{minipage}[t]{\imagewidthb}
\centering
\includegraphics[width=\imagewidthb]{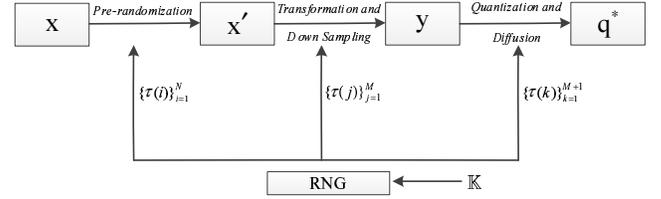}
\end{minipage}
\centering \caption{A block diagram showing the flow of compressing an encrypted image.}
\label{fig:ComEnc}
\end{figure}

As shown in Fig.~\ref{fig:ComEnc}, a plain-image $\mathbf{x}$ is \textit{pre-randomized} with $\mathbf{R}$, which can be regarded as a weak or lightweight encryption of
$\mathbf{x}$ performed by a user with limited computing power. Then a service provider performs the \textit{transformation, down sampling and joint quantization and diffusion}  processes on the ciphertext $\mathbf{x}'$ sequentially.
Referring to Sec.~\ref{sec:sampling}, it is easy to conclude that the compression ratio (CR) is exactly equal to SR.
Finally, authorized users can perform joint decompression and decryption on the quantized and diffused measurement vector $\mathbf{q}^*$ to reconstruct
an approximated version of $\mathbf{x}$.

\vspace*{-0.4\baselineskip}
\section{Conclusions}
\label{sec:conclusion}

A joint quantization and diffusion approach has been proposed for the block compressed sensing of natural images.
To realize the goal of fast compressed image sensing with secrecy,
structurally random matrix is employed for fast computation and the obtained measurements are concealed by
the proposed joint quantization and diffusion mechanism. Theoretical analyses and experimental results have demonstrated that the new scheme
can resist known-plaintext attack and its reconstruction quality is comparable to that of CS without quantization.
Our approach can be extended to compress encrypted images.


%
%



%
\bibliographystyle{IEEEbib}
\bibliography{SecureFastImaging}

\begin{thebibliography}{10}
\providecommand{\url}[1]{#1}
\csname url@samestyle\endcsname
\providecommand{\newblock}{\relax}
\providecommand{\bibinfo}[2]{#2}
\providecommand{\BIBentrySTDinterwordspacing}{\spaceskip=0pt\relax}
\providecommand{\BIBentryALTinterwordstretchfactor}{4}
\providecommand{\BIBentryALTinterwordspacing}{\spaceskip=\fontdimen2\font plus
\BIBentryALTinterwordstretchfactor\fontdimen3\font minus
  \fontdimen4\font\relax}
\providecommand{\BIBforeignlanguage}[2]{{%
\expandafter\ifx\csname l@#1\endcsname\relax
\typeout{** WARNING: IEEEtran.bst: No hyphenation pattern has been}%
\typeout{** loaded for the language `#1'. Using the pattern for}%
\typeout{** the default language instead.}%
\else
\language=\csname l@#1\endcsname
\fi
#2}}
\providecommand{\BIBdecl}{\relax}
\BIBdecl

\bibitem{Tao:BPunderRIP:TIT05}
E.~J. Candes and T.~Tao, ``Decoding by linear programming,'' \emph{IEEE Trans.
  Inf. Theory}, vol.~51, no.~12, pp. 4203--4215, Dec. 2005.

\bibitem{Donoho:CSIntro:TIT06}
D.~L. Donoho, ``Compressed sensing,'' \emph{IEEE Trans. Inf. Theory}, vol.~52,
  no.~4, pp. 1289--1306, Apr. 2006.

\bibitem{Richard:CSIntro:SPM07}
R.~Baraniuk, ``Compressive sensing,'' \emph{IEEE Signal Process. Mag.},
  vol.~24, no.~4, pp. 118--121, Jul. 2007.

\bibitem{Richard:ProveRIP:2008}
R.~Baraniuk, M.~Davenport, R.~Devore, and M.~Wakin, ``A simple proof of the
  restricted isometry property for random matrices,'' \emph{Constr. Approx.},
  vol.~28, no.~3, pp. 253--263, Dec. 2008.

\bibitem{Do:TSP:SRM12}
T.~T. Do, L.~Gan, N.~H. Nguyen, and T.~Tran, ``Fast and efficient compressive
  sensing using structurally random matrices,'' \emph{IEEE Trans. Signal
  Process.}, vol.~60, no.~1, pp. 139--154, Jan. 2012.

\bibitem{Candes:CS:TIT06}
E.~J. Candes and T.~Tao, ``Near-optimal signal recovery from random
  projections: Universal encoding strategies?'' \emph{IEEE Trans. Inf. Theory},
  vol.~52, no.~12, pp. 5406--5425, Dec. 2006.

\bibitem{Rachlin:secrecy:08}
Y.~Rachlin and D.~Baron, ``The secrecy of compressed sensing measurements,'' in
  \emph{Proc. 46th Annu. Allerton Conf. Commun. Contr. Comput.}, 2008, pp.
  813--817.

\bibitem{Bianchi:SecRLM:ICASSP14}
T.~Bianchi, V.~Bioglio, and E.~Magli, ``On the security of random linear
  measurements,'' in \emph{Proc. IEEE Int. Conf. Acoust. Speech Signal Process.
  (ICASSP)}, 2014, pp. 4020--4024.

\bibitem{Feng2014SPLQuanti}
J.-M. Feng and F.~Krahmer, ``An {RIP}-based approach to {$\Sigma \Delta$}
  quantization for compressed sensing,'' \emph{IEEE Signal Processing Letters},
  vol.~21, no.~11, pp. 1351--1355, 2014.

\bibitem{Cormen:IntroductionToAlg:09}
T.~H. Cormen, C.~E. Leiserson, R.~L. Rivest, and C.~Stein, \emph{Introduction
  to Algorithms}.\hskip 1em plus 0.5em minus 0.4em\relax MIT Press, 3rd
  edition, 2009.

\bibitem{figueiredo2007gradient}
M.~A. Figueiredo, R.~D. Nowak, and S.~J. Wright, ``Gradient projection for
  sparse reconstruction: {A}pplication to compressed sensing and other inverse
  problems,'' \emph{IEEE Journal of Selected Topics in Signal Processing},
  vol.~1, no.~4, pp. 586--597, 2007.

\bibitem{Liu14SRBitDepth}
H.~Liu, B.~Song, F.~Tian, and H.~Qin, ``Joint sampling rate and bit-depth
  optimization in compressive video sampling,'' \emph{IEEE Transactions on
  Multimedia}, vol.~16, no.~6, pp. 1549--1562, 2014.

\bibitem{kumar2009lossy}
A.~A. Kumar and A.~Makur, ``Lossy compression of encrypted image by compressive
  sensing technique,'' in \emph{Proc. of IEEE Region 10 Conf.}, 2009, pp. 1--5.

\bibitem{zhang2011compressing}
X.~Zhang, Y.~Ren, G.~Feng, and Z.~Qian, ``Compressing encrypted image using
  compressive sensing,'' in \emph{Proc. 7th Int. Conf. on Intelligent
  Information Hiding and Multimedia Signal Process. (IIH-MSP)}, 2011, pp.
  222--225.

\end{thebibliography}

\end{document}